# Exploring a CNN Model for Earthquake Magnitude Estimation using HR-GNSS data

Claudia Quinteros-Cartaya[1], Jonas Köhler[1,2], Wei Li[1], Johannes Faber[1,3], & Nishtha Srivastava[1,2*]

1. Frankfurt Institute for Advanced Studies, Germany.

2. Institute of Geosciences, Goethe - University Frankfurt, Germany.

3. Institute for Theoretical Physics, Goethe – University Frankfurt, Germany.

* Corresponding Author e-mail: srivastava@fias.uni-frankfurt.de

## Abstract

High-rate Global Navigation Satellite System (HR-GNSS) data can be highly useful for earthquake analysis as it provides continuous high-rate measurements of ground motion. This data can be used to estimate the magnitude, to assess the potential of an earthquake for generating tsunamis, and to analyze diverse parameters related to the seismic source. Particularly, in this work, we present the first results of a deep learning model based on a convolutional neural network for earthquake magnitude estimation, using HR-GNSS displacement time series. The influence of different dataset configurations, such as station numbers, epicentral distances, signal duration, and earthquake size, were analyzed to figure out how the model can be adapted to various scenarios. We explored the potential of the model for global application and compared its performance using both synthetic and real data from different seismogenic regions. The performance of our model at this stage was satisfactory in estimating earthquake magnitude from synthetic data with $0.07 \leq \text{RMS} \leq 0.11$. Comparable results were observed in tests using synthetic data from a different tectonic region than the training data, with $\text{RMS} \leq 0.15$. Furthermore, the model was tested using real data from different regions and magnitudes, resulting in a good accuracy, in the best cases with $0.09 \leq \text{RMS} \leq 0.33$, provided that the data from a particular group of stations had similar epicentral distance constraints to those used during the model training. The robustness of the DL model can be improved to work independently from the window size of the time series and the number of stations, enabling faster estimation by the model using only near-field data. Overall, this study provides insights for the development of future DL approaches for earthquake magnitude estimation with HR-GNSS data, emphasizing the importance of proper handling and careful data selection for further model improvements.

## 1. Introduction

The Global Navigation Satellite System (HR-GNSS) can provide high-frequency and high-precision position measurements that facilitate the detection of very small ground displacements caused by large earthquakes. Unlike inertial sensors, HR-GNSS instruments can record the signal of large earthquakes near the source without saturation, providing valuable information on both dynamic (far-field) and static (near-field) displacements (Bock et al., 2000; Ge et al., 2000; Kouba, 2003; Larson, 2009).

Furthermore, earthquake magnitude estimation from GNSS waveforms has been made possible through empirical relationships between the peak ground displacement and the seismic moment (Crowell et al., 2013; Melgar et al., 2015; Goldberg et al., 2021). In the last decades, researchers have explored incorporating GNSS data to improve the accuracy of earthquake magnitude estimation compared to using seismic data alone from seismometers and accelerometers (Bock et al., 2011; Wang et al., 2013). As a result, HR-GNSS networks for earthquake monitoring and the continuous recording of data for near-real-time analysis have increased.

The 2004 Sumatra-Adaman earthquake with a magnitude of Mw 9.1 was a significant event that motivated the implementation and improvement of early warning systems in potential seismogenic regions using HR-GNSS sensors (Blewitt et al., 2006; Satake, 2014). Subsequent great earthquakes such as the Mw 8.8 Maule (2010), Mw 9.0 Tohoku (2011), and Mw 8.4 Illapel (2015), among others, have demonstrated the importance of the fast and reliable assessment of seismic sources, leading to the development of diverse algorithms for HR-GNSS data analysis that enable proper early warning of earthquakes and tsunamis (e.g., Crowell et al., 2009; 2016; 2018; Allen & Ziv, 2011; Fang et al., 2014; Grapenthin et al., 2014; Minson et al., 2014; Kawamoto et al., 2016; Ruhl et al., 2017; 2019; Psimoulis et al., 2018).

Moreover, seismologists aim to develop complementary tools to outperform traditional analysis methods through deep learning (DL) approaches, which have proven to have great capacity in big data processing and feature extraction for fast and robust results. DL methods have been widely introduced to deal with various seismological tasks, such as earthquake detection, phase picking, seismic source assessment, and denoising of seismic signals (e.g., Chakraborty, et al., 2022a; 2022b; Jiao & Alavi, 2020; Kuang et al., 2021; Li, 2022a; Li, 2022b; Mousavi & Beroza, 2022; Perol et al., 2018; van den Ende et al., 2020). However, training DL algorithms with HR-GNSS data for seismic analysis is one of the most recent challenges still in development. For instance, Lin et al. (2021) showed a first demonstration of seismic source patterns analysis and magnitude estimation through a DL algorithm and HR-GNSS data, which focused on the seismic activity in the Chile subduction zone by using peak ground displacement time series. On the other hand, Dittmann et al. (2022) introduced a DL algorithm for earthquake detection through velocity time series obtained from HR-GNSS by the time-differenced carrier phase.

In this work, we present a preliminary DL model based on a convolutional neural network for the magnitude estimation of HR-GNSS data. Unlike previous algorithms, our model is trained merely on displacement time series, and we evaluate the possibility of extending its application on a global scale. We tested the performance of our model using both synthetic and real data from different seismogenic regions. We also analyzed the influence of different dataset configurations, such as epicentral distances, signal duration, and earthquake size, to determine how the model can be adapted to different scenarios.

## 2. Architecture

We propose a deep learning model using a sequential Convolutional Neural Network (CNN) for a regression problem (Le Cun, 1989; Schmidhuber, 2015; Géron, 2019; Goodfellow et al., 2016). The CNN architecture consists of six 2D-convolutional layers, three max-pooling layers (Scherer et al., 2010; Zhou & Chellappa, 1988), and three fully-connected or dense layers (Le Cun et al., 1998).

The input layer for the DL model is comprised of displacement time series for each earthquake with 1 Hz sampling rate. These time series are stored in a tensor with dimensions $N_s$ x $N_t$ x 3, where $N_s$ represents the number of stations, $N_t$ represents the number of samples in the time series, and three channels correspond to the U, N, and E components referring to the up, north, and east directions of the sensor in each GNSS station (as shown in Figure 1).

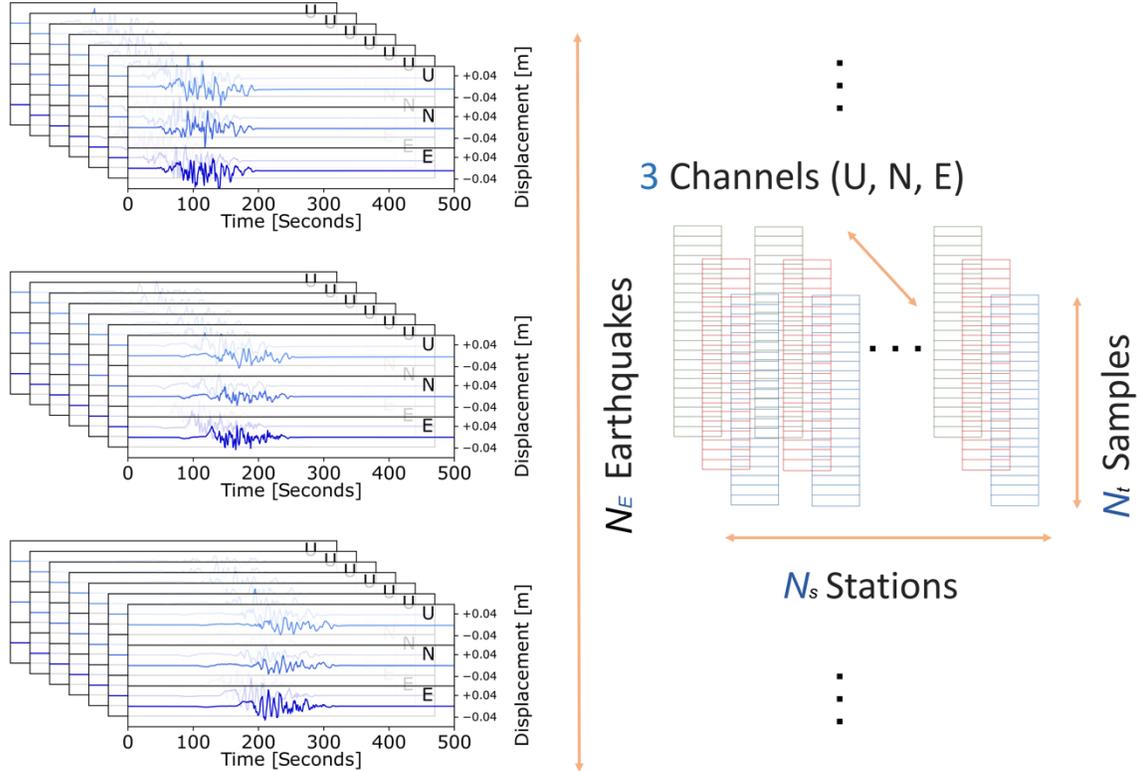

**Figure 1.** The input data for the HR-GNSS displacement time series is stored in a tensor whose shape depends on the number of earthquakes ($N_E$), station numbers ($N_s$), samples in the time series ($N_t$), and 3 channels (U: up, N: north, and E: east directions). The amplitudes in the time series represent displacements in meters, and the sampling rate is 1Hz, with every sample representing one second in time.

The architecture of our model is summarized in Figure 2. For each convolutional layer, we used different numbers of filters with kernel size (1, 3) and stride (1, 1). A zero-padding was only used in the first and the last convolutional layer. A pool size of (1, 2) was used in each max-pooling layer. Thus, we are down sampling the data, while keeping the extraction of features in the time series separated by station up to the dense layers.

We chose a rectified linear unit activation function (ReLU) as the transfer function to activate the output in every convolutional and dense layer (Nair & Hinton, 2010). The last tensor that results from the convolutional layers is transformed through a flattening layer to a one-dimensional vector (Krizhevsky et al., 2017) to be the input for the dense layers.

Then, the three dense layers consist of 128, 32, and 1 neuron, respectively. The weights of the kernels for the two first dense layers are initialized using a normal distribution and constrained by max-norm regularization with a maximum norm value of 3 (Géron, 2019).

Since we do not adopt any normalization for the values of the labels in the training (magnitudes), we obtain a target variable in the output layer whose value is equivalent to the earthquake moment magnitude Mw (Hanks & Kanamori, 1979).

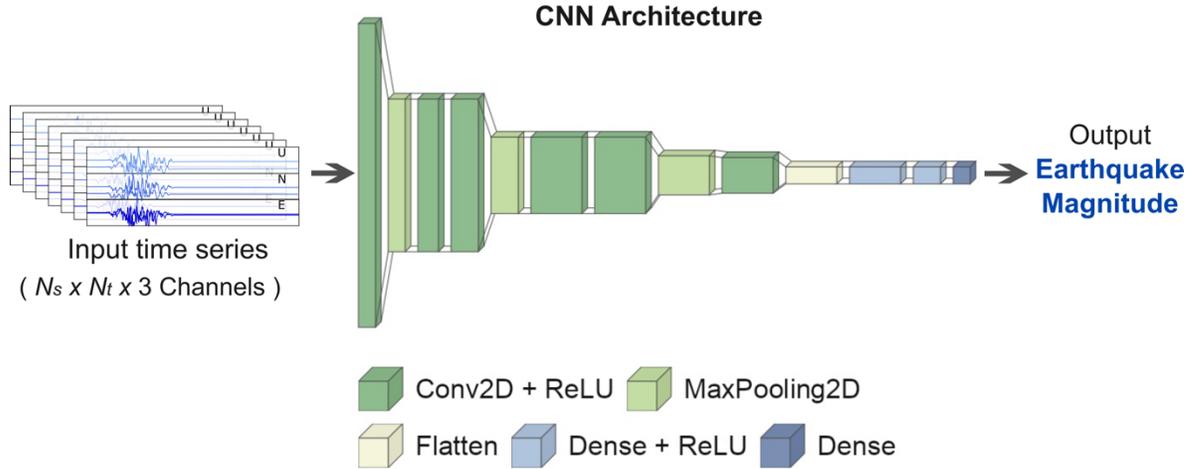

**Figure 2.** Sequential Convolutional Neural Network architecture proposed in this work for earthquake magnitude estimation using displacement time series in three components (3 channels) from a specific number of HR-GNSS stations, $N_s$, and number of samples, $N_t$, in the time domain.

## 3. Data

In optimal cases, HR-GNSS stations located within about 5 km from the epicenter can detect displacements caused by moderate earthquakes with a magnitude of around Mw 5 (e.g., Mendoza et al., 2013). However, to increase the likelihood of having sufficient HR-GNSS recordings to observe earthquake signals, we focused our analysis on Mw > 6 earthquakes. Nonetheless, the number of large earthquakes available for analysis using HR-GNSS instruments may not be enough to form a representative dataset for training a deep learning (DL) model.

Therefore, we utilized synthetic HR-GNSS signals from a previously generated database by Lin et al. (2020), containing a large volume of data of 36,800 earthquakes ranging from magnitudes of 6.6 to 9.6. These earthquakes are associated with rupture scenarios specifically modeled for the Chile subduction zone (Figure 3).

We utilized the synthetic HR-GNSS data from the Chile region for training, validation, and testing of our DL model. Furthermore, we evaluated the performance of our model by testing it with synthetic signals of a Mw 8.7 Cascadia earthquake (Melgar et al., 2016) and real data (Melgar & Ruhl, 2018) from six large earthquakes from diverse regions (Figure 4).

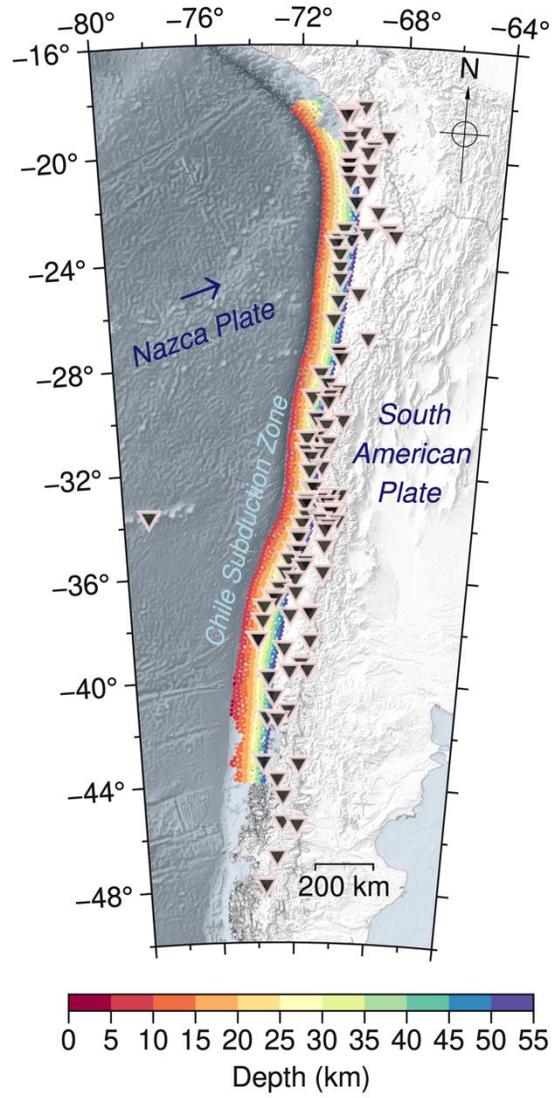

**Figure 3.** Chile Subduction Zone. The synthetic data represent the displacements hypothetically recorded by the HR-GNSS stations shown as black triangles on the map (Báez et al., 2018), corresponding to the earthquakes whose hypocenters are shown as dots in color scale by depth.

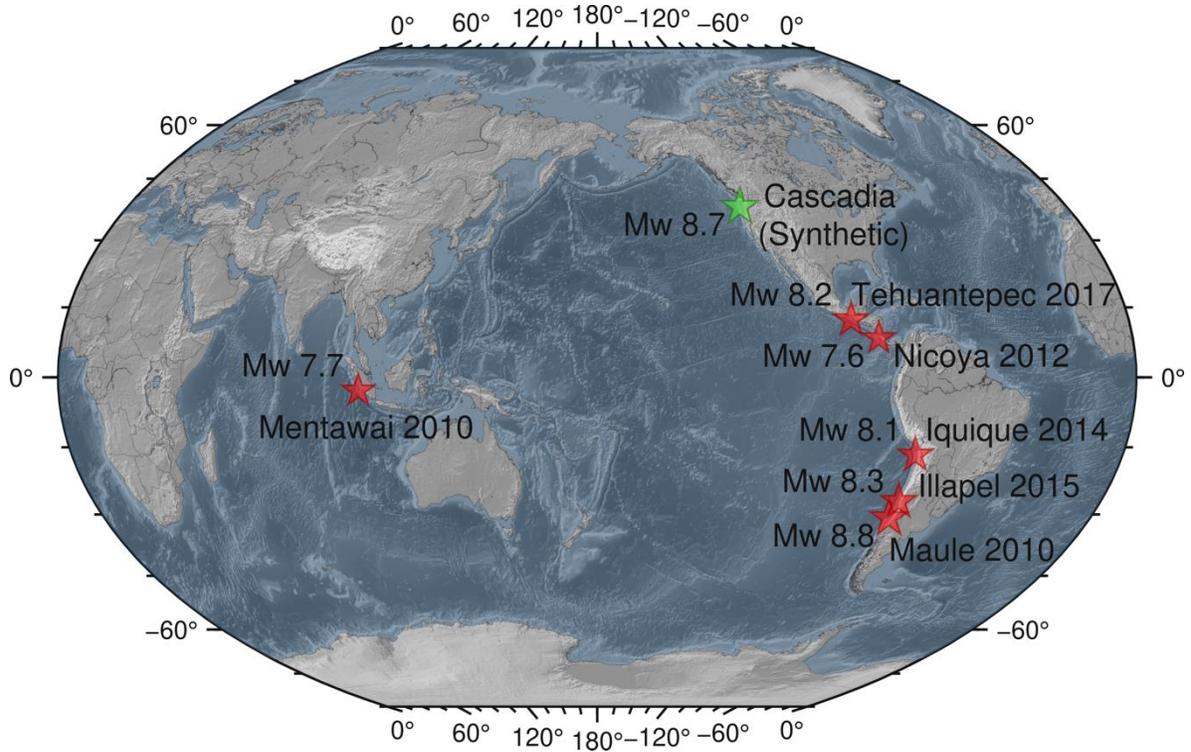

**Figure 4.** Earthquakes used for testing the model. Red stars represent the epicenter of the earthquakes used with real HR-GNSS signals, and the green star is the epicenter of the Cascadia earthquake used with synthetic signals.

## 4. Training

*4.1 Data preparation*

Since we explore the influence of using data from different numbers of stations, epicentral distances, and time series lengths on the model performance, we prepared data sets to correspond to three cases: Case I, Case II, and Case III, which involved the training of three models with the same architecture, but using different shapes for the input data (Table 1). The initial time in the time series, t=0, was referenced to the earthquake origin time. The amplitude displacements have physical units in meters, and the sampling is in the time domain, in seconds.

For Case I and Case II, we used data from three and seven stations, respectively, located within a 3º radius from the epicenter ($\Delta \leq 3º$; Figure 5a), and time series that contain 180 seconds after the earthquake origin time. Then, for Case III, we sought to incorporate a greater range of maximum displacements observed in the time series, particularly those in the near and mid-field (Blewitt et al., 2006). To achieve this, we utilized data from seven different stations (Figure 5b), ensuring a balanced representation by selecting at least three and up to five stations within a radius of 3º from the epicenter. For the remaining stations farther than 3º away, we included those with epicentral distances that did not cause amplitude displacements too small to detect in the time series. The maximum distance depended on the magnitude of the event and how the modeling was previously constrained to generate the synthetic dataset (Lin et al., 2021; Lin et al., 2020). In this last case, Case III, every time series contains the first 500 seconds after the origin time of the earthquakes.

The stations were selected randomly for each case, with the caveat that we avoided having data from several stations too close to each other for a particular earthquake. We included only those cases in which at least three stations had azimuths that differed by 40º from each other. Thus, we aimed to have time series with features as different as possible, such as amplitude values, time wave arrivals, duration of the earthquake signal, and so on. Lastly, we selected a total of 34,567 earthquakes and split them into different sets: 90% for the learning process (training and validation) and 10% for testing.

**Table 1.** Setting of the input data for the three cases. Each case corresponds to one training instance of the same architecture for the DL models, but the input data have shapes that differ between the cases.

| Cases | Number of Stations per earthquake (Δ: epicentral distance) | Time series window size (samples) | Input Data Shape |
|---|---|---|---|
| Case I | 3 stations, Δ ≤ 3º | 181 | (3 x 181 x 3*) |
| Case II | 7 stations, Δ ≤ 3º | 181 | (7 x 181 x 3*) |
| Case III | 7 stations: from 3 to 5 stations Δ ≤ 3º and the rest Δ > 3º | 501 | (7 x 501 x 3*) |

* 3 Channels: components in U, E, and N directions.

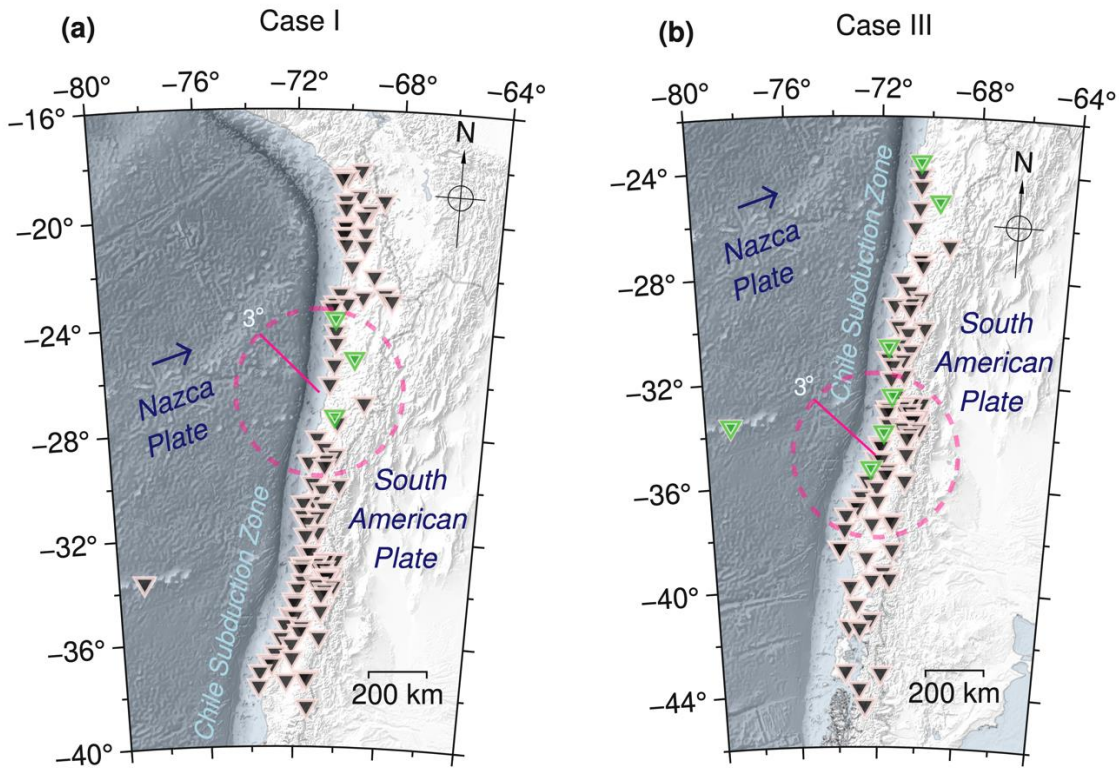

**Figure 5.** Example of station distributions for the selection of data. The earthquake epicenter is in the center of the circle and the stations that were randomly chosen are shown as green triangles. In (a), three stations with epicentral distances Δ ≤ 3º are shown as an example of station distributions for Case I. For Case II, we could use the same example as Case I, but choose seven stations instead. In (b), a particular example of seven stations for Case III: three stations Δ ≤ 3º, and four stations Δ > 3º.

*4.2 Learning process*

We split the synthetic dataset into a training set and a validation set, taking 90% of the earthquakes for the learning process (Figure 6). The training set was further split into an 80/20 ratio, with 80% used for training (72% of the total earthquakes in the database) and 20% for fine-tuning hyperparameters that control the learning through the validation process (18% of the total earthquakes in the database).

Each earthquake in the training set was labeled with a target variable that corresponds to its magnitude value rounded to one decimal. The mean squared error function (MSE) was used to evaluate the losses during the training process. We optimized the model using the Adaptive Moment (ADAM) estimation method to reduce the losses (Géron, 2019; Kingma & Ba, 2015). To prevent update steps from exceeding the initial learning rate, we used a learning rate schedule with a standard decay function (decay rate = learning rate/epochs). This helped to increase the performance of the training. We set the initial learning rate to 0.01, the decay to 0.1/maximum number of epochs (see Appendix A), the maximum number of epochs to 200, and the batch size to 128. We used early stopping to reduce the possibility of overfitting, stopping the training process when the minimum validation loss was reached, with a patience of 20 epochs.

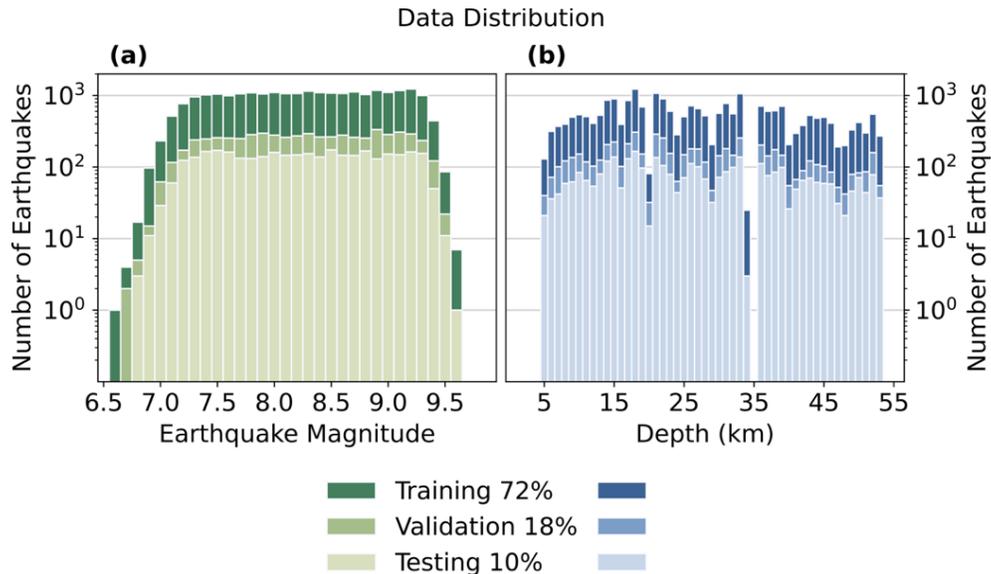

**Figure 6.** Histograms of earthquake distribution in the database are shown, separated by magnitude (left) and depth (right). The dataset was split into three sets: 72% for training, 18% for validation during training, and 10% for testing. The number of earthquakes is displayed on a logarithmic scale to better visualize the smaller data groups.

## 5. Testing

*5.1 Chile Synthetic data*

We evaluated the performance of the three models using 10% of the earthquakes in the database that were not used for training or validation (Figure 6). The selection criteria for the testing set were the same as those for the training set in each case.

As shown in Figure 7, the magnitude estimations resulted in low errors in all three cases. The lowest root mean squared error (RMS) of 0.06 was achieved in Case III, where most of the estimated magnitudes were accurate. The error distributions by magnitude are also presented in Figure 7d, 7e, and 7f.

In particular, for Case I and Case II, the majority of the estimated magnitudes in the range of $7.0 \leq Mw \leq 8.3$ were accurate. For Case III, the best fit was observed for almost all magnitudes, ranging from Mw 7.0 to 9.6. In all three cases, the errors increased with increasing magnitude. This trend was more noticeable in Case I and Case II, where the RMS values increased from Mw 7.9. In contrast, the RMS values slightly increased from Mw 9.2 in Case III.

For lower magnitudes (Mw ≤ 6.9), the estimations were mostly overestimated in all three cases. For higher magnitudes (Mw ≥ 9.5), the estimations were underestimated in Case I and Case II. However, due to the uneven distribution of earthquakes by magnitude in the testing set, we had very few earthquakes with Mw ≤ 6.9 and Mw ≥ 9.5, resulting in a higher error for those magnitudes and an RMS value that may not be as representative as those of other magnitudes.

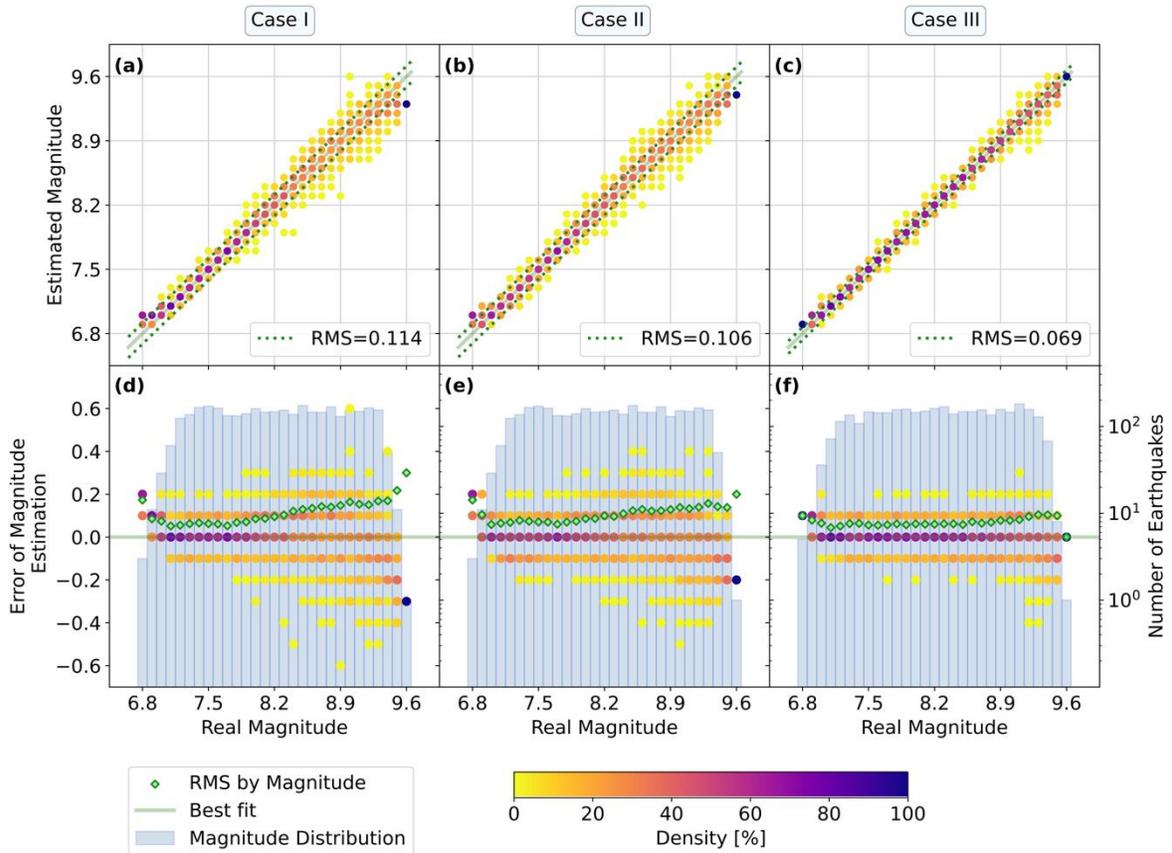

**Figure 7.** Error of magnitude estimation using synthetic HR-GNSS data from Chile, for every case described in Table 1. Each circle corresponds to one bin whose color represents the percentage of tests done for each real magnitude. Plots (a), (b), and (c) are the fits of the magnitude estimations, where the RMSs are shown by the dotted lines. The results are binned on a 0.1 magnitude grid. Plots (d), (e), and (f) are the distribution of errors and RMS by each magnitude. The errors correspond to the difference between the real magnitude and the magnitude estimated by the model. The results are binned on a grid, 0.1 magnitude × 0.1 magnitude error. The RMS values by each magnitude are shown as green diamond symbols. The testing data distribution by magnitudes is represented through histograms in the background of the plots.

In the case of the highest magnitudes, the estimations may be affected by the windowing of the time series. For the largest earthquakes, 181 seconds in the recordings from stations nearly 3º away from the epicenter may not be long enough to contain the complete earthquake signal, resulting in increased error in Case I and Case II. Figure 8 shows a comparison of waveforms from earthquakes with different magnitudes and epicentral distances, in displacement time series of 181 and 501 seconds. For example, the synthetic waveforms of an Mw 6.7 earthquake with epicentral distances Δ ≤ 3º are complete before 150 seconds, whereas the waveforms of an Mw 8.1 earthquake, from stations Δ ≤ 3º, just barely fit within a time window of 181 seconds. Furthermore, the waveforms of an Mw 9.0 earthquake require more than 181 seconds to fit the complete signal for epicentral distances larger than 1.5º.

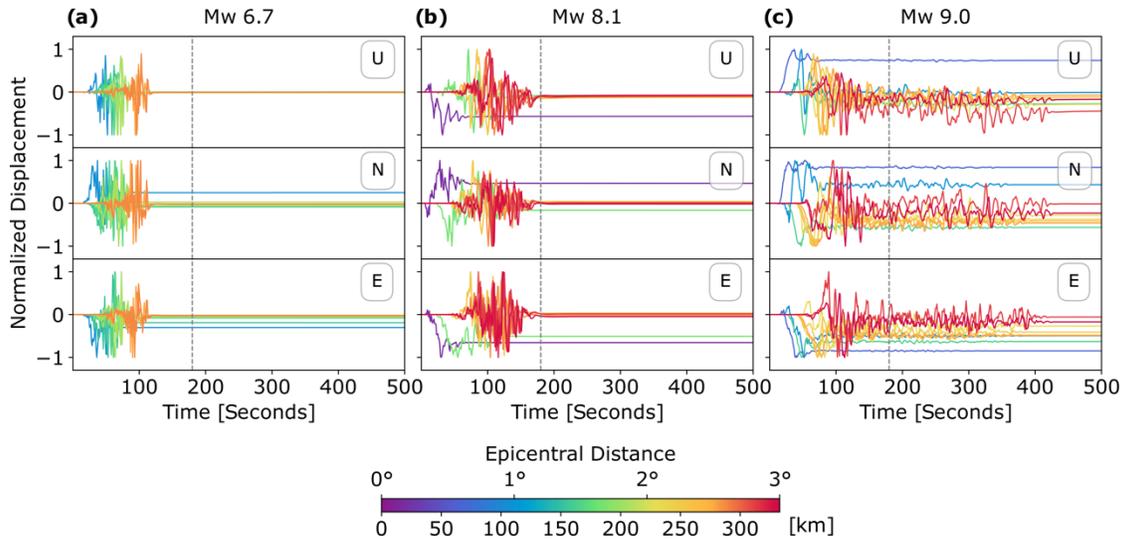

**Figure 8**. Comparison of HR-GNSS synthetic data from earthquakes of magnitudes Mw 6.7, 8.1, and 9.0, in three components (U: up, N: north, and E: east directions). The initial time is referenced to the origin time of the earthquake. The amplitudes are normalized by their maximum absolute value and colors are related to the epicentral distance of stations (for distances Δ ≤ 3º). The dashed lines refer to a window size of 181 seconds. Synthetic data provided by Lin et al., 2020)

*5.2 Comparison using Chile and Cascadia Synthetic Data*

To evaluate the performance of the models using synthetic data from regions with different tectonic regimes, we tested the models with two Mw 8.7 earthquakes: one from Chile (Lin et al., 2020) and the other from Cascadia (Melgar et al., 2016). We randomly selected an Mw 8.7 earthquake from the testing dataset of Chile. The data corresponded to displacement time series from 63 HR-GNSS stations for the Chile earthquake and 62 HR-GNSS stations for the Cascadia earthquake. For each earthquake, we randomly generated 500 distinct groups of three and seven stations, with a window size of the time series according to each case.

In Figure 9, the results of the models in the three cases are quite similar for both earthquakes. Although the results of the model in Case III are again the best in most estimations, with the least scattered errors and an RMS of 0.07 for Chile and 0.11 for Cascadia, in general, all three cases show a suitable performance of the model with a relatively low RMS.

There is no discernible pattern indicating whether those groups of stations farther away from the epicenter tend to have a higher error than the group with the closest stations to the epicenter. However, as we outlined in the first testing (section 5.1), for great magnitudes, the performance of the model is better when using time series with long window sizes. Hence, we assume that the use of time series long enough in Case III could contribute to obtaining better results than in the other two cases.

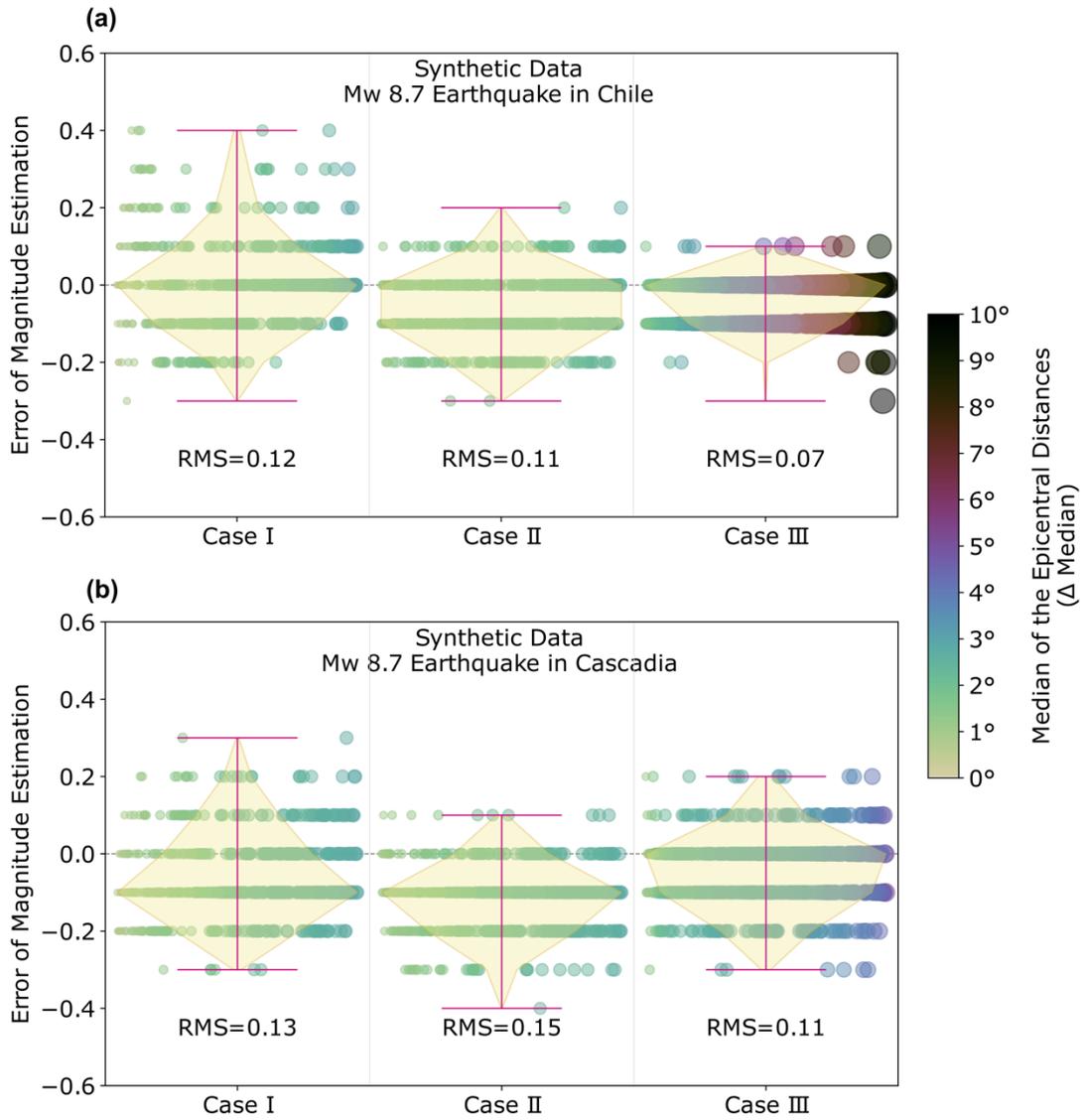

**Figure 9.** Errors of magnitude estimations using synthetic data from an Mw 8.7 Cascadia and an Mw 8.7 Chile earthquake, for the DL models in the three cases (Table 1). The circles represent the magnitude error for each group of stations defined by 500 different random combinations. Both the color scale and circle sizes (sorted from left to right) depend on the median of the epicentral distances (Δ Median) of each combination of stations, in each case.

*5.3 Real earthquakes*

The noise in real GNSS signals is a significant factor that could affect the accuracy of magnitude estimations, especially if the DL model used was trained with synthetic data that lacks noise. Therefore, we tested the models when faced with real HR-GNSS data from six earthquakes with different magnitudes and from different tectonic regions (Figure 4): Maule, Iquique, and Illapel earthquakes, from Chile, and the others from Tehuantepec (Mexico), Nicoya (Costa Rica), and Mentawai (Indonesia).

The waveforms of these earthquakes are from the database provided by Melgar & Ruhl (2018) and consist of displacement waveforms with a signal-to-noise ratio greater than 3 dB and a minimum peak amplitude of 0.04 m (Ruhl et al., 2019). Because the number of stations in the database differs for each earthquake, we tested with different numbers of station groups for each one. Thus, we had some cases with only one group tested (Nicoya, Tehuantepec, and Mentawai), whereas for the Illapel, Maule, and Iquique, we had numerous station groups, from which we randomly selected 500 cases for testing.

In Figure 10, we show the results for every case and earthquake. In cases such as Iquique and Tehuantepec, the magnitude estimations reached RMS values of 0.09 and 0.1, respectively, which are comparable to the tests done using synthetic data. However, in general, we observed more scattered error values in these results from real data than from synthetic data. Our first assumption is that the noise content in real data causes an increase in the error value since our models are trained with ideal and clean data. However, we can analyze some other considerations:

1. The highest RMS value of 0.49 was obtained for the Mw 7.6 Nicoya earthquake. This occurred when all stations in the group were within an epicentral distance of less than 1º, and large amplitudes, with displacements of nearly 50 cm, were observed in the first 20 seconds (Figure 11a). It is possible that the model overestimated the magnitude in this case because, during training, the grouped stations were rarely within such a short radius with large displacements.
2. The results for the Mw 7.7 Mentawai earthquake are acceptable for Case I and Case II, with an RMS of 0.2. However, in Case III, the RMS of 0.44 is associated with the same cause as the Nicoya earthquake since, in this case, only one station is at $\Delta > 3º$ (Figure 11b), and the model was trained using groups with at least two stations $\Delta > 3º$.
3. As mentioned above, the Mw 8.1 Iquique earthquake had the best fit in Case II, with an RMS value of 0.09. But for Case III, the RMS value was 0.33, obtained from groups of stations with epicentral distances whose median is less than 3.5º. Despite having data from stations with epicentral distances of up to approximately 8º, the dataset only contained three stations farther than 3.5º away (Figure 12a).
4. The most accurate magnitude estimations for the three largest earthquakes: Tehuantepec, Illapel, and Maule, were obtained in Case III, with RMS values of 0.1, 0.17, and 0.13, respectively. These results are consistent with the results obtained from synthetic data. Specifically, the test using data from the Mw 8.2 Tehuantepec earthquake only had seven stations, and only two of them were located within a 3º radius from the epicenter (Figure 11c). In Cases I and II, stations located more than 3º away were used in the models, which resulted in waveforms with smaller displacements than those used in the model training. Therefore, the magnitude estimations for these models were expected to be underestimated. However, the errors for these models were still low.
5. For the Mw 8.8 Maule earthquake, only five stations were located within a radius of 3º from the epicenter (Figure 12c). In Case II, stations located at $\Delta > 3º$ were used in the testing. This could have caused the scattered errors observed in Case II and higher errors than those in Case I and Case III.

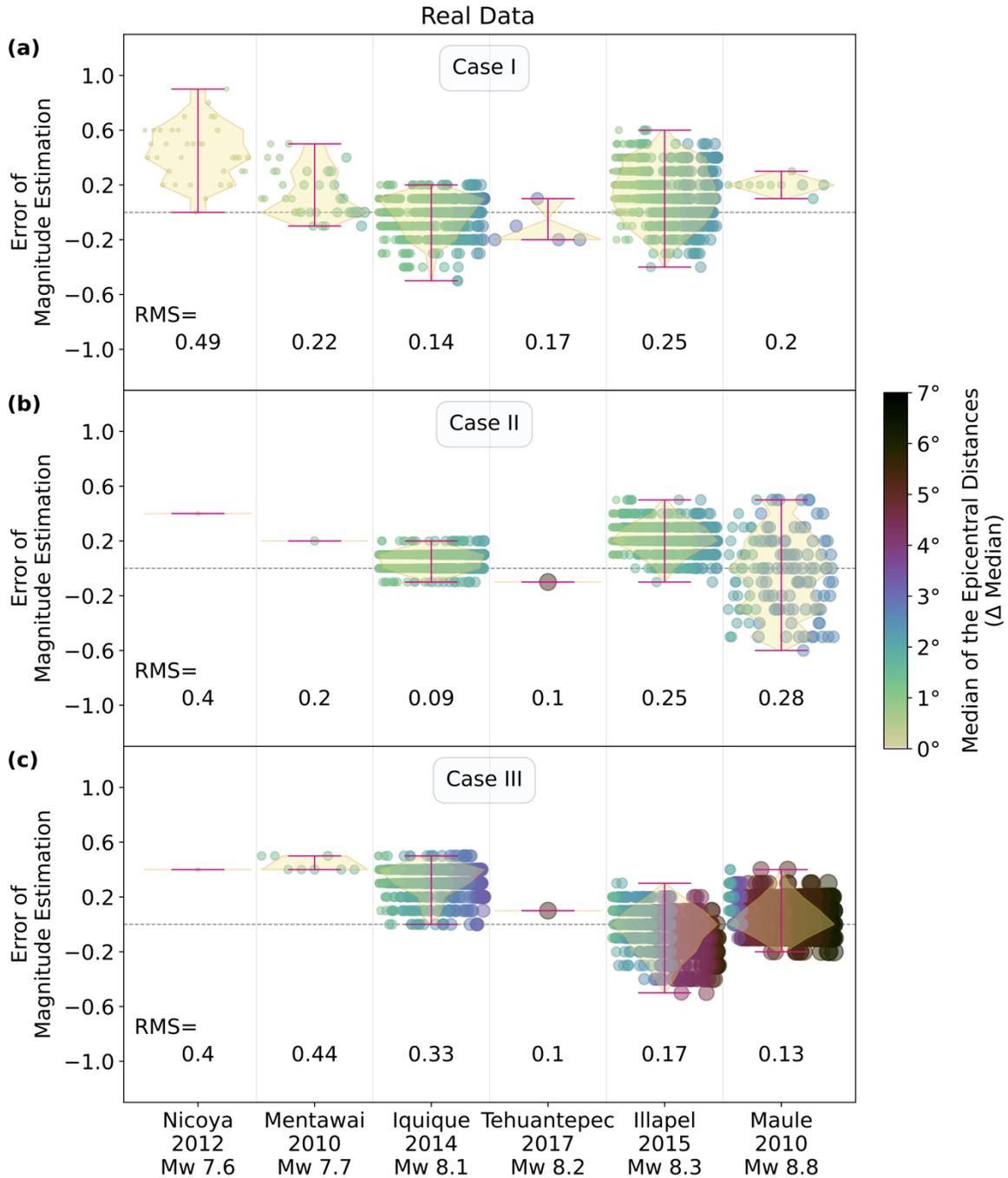

**Figure 10.** Errors of magnitude estimations using real data of earthquakes from different regions and with different magnitudes. The plots correspond to the results in Case I, Case II, and Case III, from up to the bottom, respectively. The circles indicate the magnitude error for each group of stations, which were defined by different random combinations. Both the color scale and circle sizes (sorted from left to right) depend on the median of the epicentral distances (Δ Median) of each combination of stations, for each earthquake.

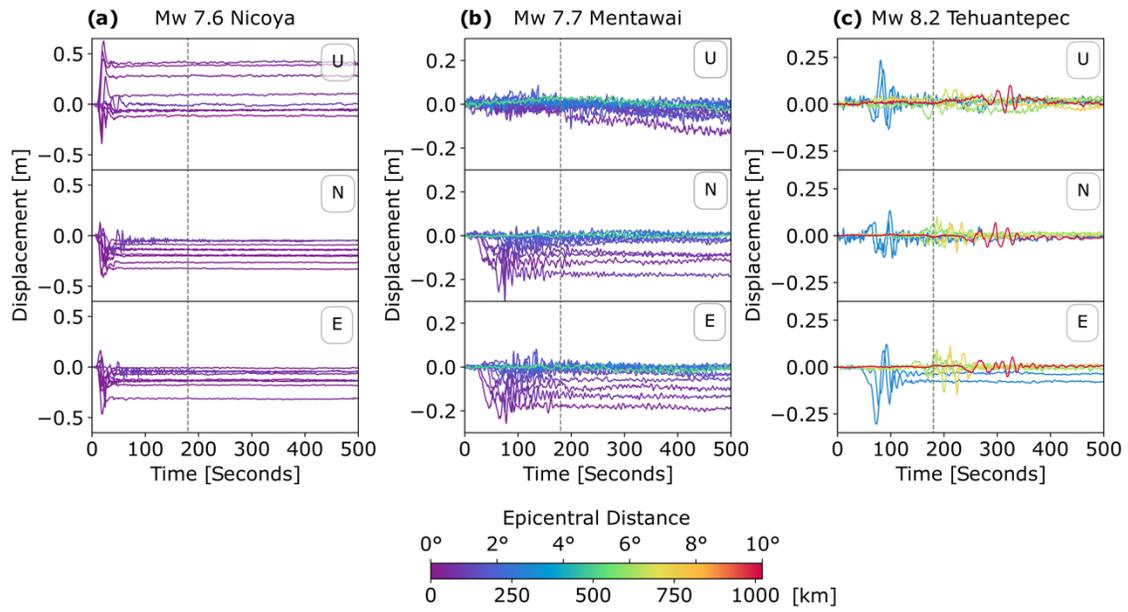

**Figure 11.** HR-GNSS real data from earthquakes from Tehuantepec (Mexico), Nicoya (Costa Rica), and Mentawai (Indonesia), which were used for the model testing. The time series are in three components (U: up, N: north, and E: east directions). The initial time is referenced to the origin time of the earthquake. The color scale is related to the epicentral distance. The dashed lines refer to a window size of 181 seconds.

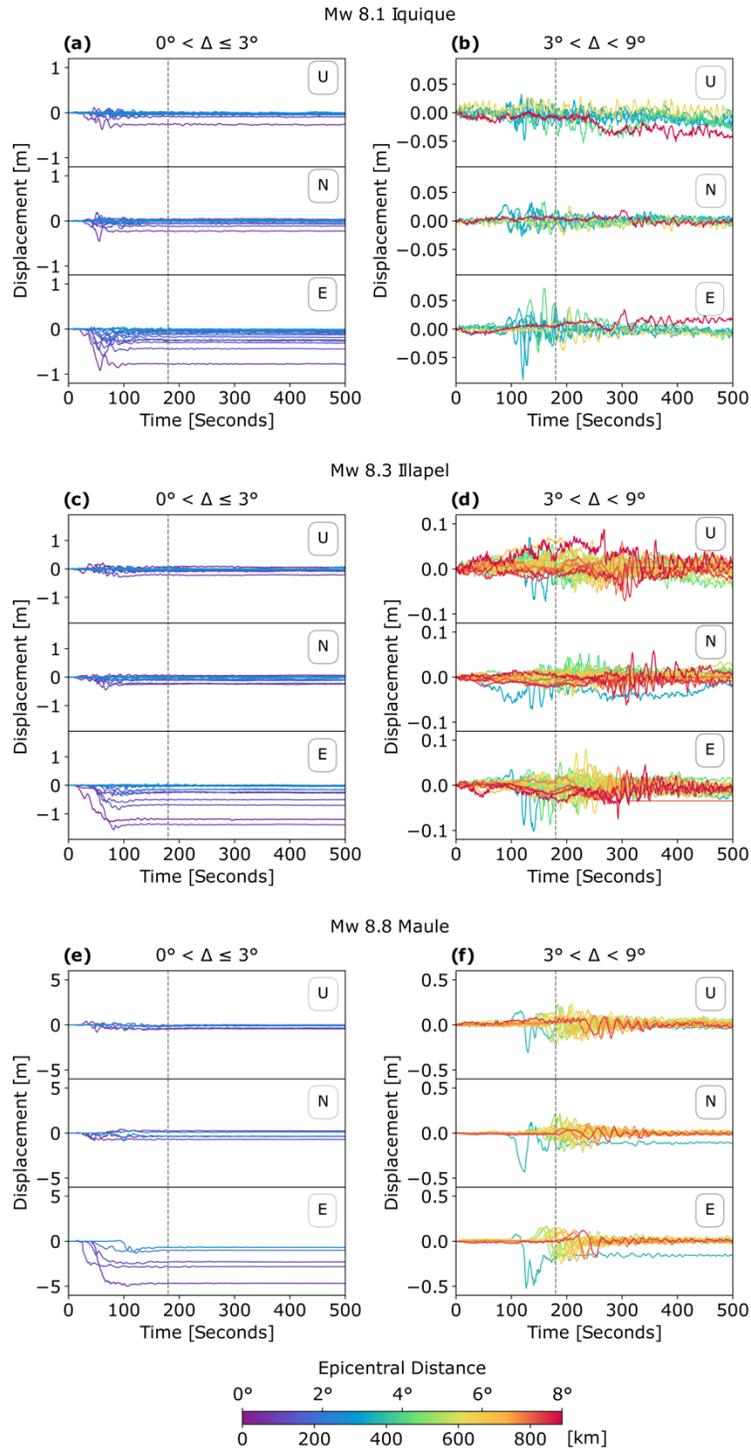

**Figure 12.** HR-GNSS real data from Maule, Iquique, and Illapel earthquakes (Chile), which were used for the model testing. The time series are in three components (U: up, N: north, and E: east directions). The initial time is referenced to the origin time of the earthquake. In the left side, the signals are from epicentral distance Δ ≤ 3º, and in the right, from Δ > 3º. Colors are related to the epicentral distance of the station. The dashed lines refer to a window size of 181 seconds.

# 6. Conclusions

The DL architecture proposed in this work is an experimental version for earthquake magnitude estimation. It has been trained using synthetic displacement time series from groups of three and seven high-rate Global Navigation Satellite System (HR-GNSS) stations, and different window sizes containing 180 samples (3 minutes) and 500 samples (over 8 minutes) after the earthquake origin time. The performance of the DL model for the estimation of earthquake magnitude from synthetic data has been satisfactory. Despite being trained with synthetic data from Chile, the model has given comparable results in tests using synthetic data from Cascadia, which represents a different tectonic region. Additionally, the results of using real data from earthquakes with different magnitudes and from different regions showed good accuracy of the estimations, provided that the data from a particular group of stations have similar epicentral distance constraints to those used during the model training. The length of the time series should also be long enough to fit most of the earthquake signals within the time window, as incomplete signals could affect the estimations.

While the DL models performed well using real data, regardless of the tectonic region of the earthquake, it would be advisable to evaluate their robustness by including noise in the training data and addressing accuracy problems related to the imbalanced amount of data by earthquake magnitude and non-normalized displacement time series in the training. This approach proposes a DL model designed for specific shapes of input data (number of stations and windowing), but the architecture could be improved to work independently from the window size of the time series and the number of stations. This would enable a faster estimation by the model using only near-field data from stations within a radius of less than 1º from the epicenter, which could provide reliable magnitude estimation with less than two minutes of data after the earthquake origin time.

The DL architecture proposed in this work is the result of preliminary analysis that will be helpful for the improvement of future DL approaches that could be used for earthquake magnitude estimation with HR-GNSS data. It is still necessary to introduce some changes in the DL model to adapt it for real-time magnitude estimation. Also, it is important to note that this is a numerical method, and therefore the robustness of the models and the physical sense depends on proper handling and selection of the training data set.

## Acknowledgments

We would like to express our sincere thanks to J. Alejandro González, Carlos Reinoza, and B. Crowell for the constructive comments and information about Geodesic Data analysis for seismology. The databases used in this work are provided https://doi.org/10.5281/zenodo.4008690 and https://doi.org/10.5281/zenodo.1434374. This research was supported by the Federal Ministry of Education and Research of Germany (BMBF), grant SAI 01IS20059. Modeling and data processing were performed at the Frankfurt Institute for Advanced Studies, with a GPU cluster funded by BMBF for the project Seismologie und Artifizielle Intelligenz (SAI).

## Statements & Declarations

*Competing Interests*

The authors have no relevant financial or non-financial interests to disclose.

*Author Contributions*

All authors contributed to the study conception and design. Conceptualization, data collection, programming, and analysis were performed by Claudia Quinteros Cartaya. Methodology was contributions by Wei li, Jonas Köhler, and Johannes Faber. Conceptualization and review by Nishtha Srivastava. The first draft of the manuscript was written by Claudia Quinteros Cartaya, and all authors commented on previous versions of the manuscript. All authors read and approved the final manuscript.

# References

Allen, R. M., & Ziv, A. (2011). Application of real-time GPS to earthquake early warning. Geophysical Research Letters, 38(16). https://doi.org/10.1029/2011GL047947

Báez, J. C., Leyton, F., Troncoso, C., Del Campo, F., Bevis, M., Vigny, C., Moreno, M., Simons, M., Kendrick, E., Parra, H., & Blume, F. (2018). The Chilean GNSS network: Current status and progress toward early warning applications. Seismological Research Letters, 89(4), 1546–1554. https://doi.org/10.1785/0220180011

Blewitt, G., Kreemer, C., Hammond, W. C., Plag, H. P., Stein, S., & Okal, E. (2006). Rapid determination of earthquake magnitude using GPS for tsunami warning systems. Geophysical Research Letters, 33(11). https://doi.org/10.1029/2006GL026145

Bock, Y., Melgar, D., & Crowell, B. W. (2011). Real-time strong-motion broadband displacements from collocated GPS and accelerometers. Bulletin of the Seismological Society of America, 101(6), 2904–2925. https://doi.org/10.1785/0120110007

Bock, Y., Nikolaidis, R. M., De Jonge, P. J., & Bevis, M. (2000). Instantaneous geodetic positioning at medium distances with the Global Positioning System. Journal of Geophysical Research: Solid Earth, 105(B12), 28223–28253. https://doi.org/10.1029/2000jb900268

Chakraborty, M., Quinteros Cartaya, C., Li, W., Faber, J., Rümpker, G., Stoecker, H., & Srivastava, N. (2022a). PolarCAP – A deep learning approach for first motion polarity classification of earthquake waveforms. Artificial Intelligence in Geosciences, 3, 46–52. https://doi.org/10.1016/j.aiig.2022.08.001

Chakraborty, M., Li, W., Faber, J., Rümpker, G., Stoecker, H., & Srivastava, N. (2022b). A study on the effect of input data length on a deep-learning-based magnitude classifier. Solid Earth, 13(11), 1721–1729. https://doi.org/10.5194/se-13-1721-2022

Crowell, B. W., Bock, Y., & Squibb, M. B. (2009). Demonstration of Earthquake Early Warning Using Total Displacement Waveforms from Real-time GPS Networks. Seismological Research Letters, 80(5), 772–782. https://doi.org/10.1785/gssrl.80.5.772

Crowell, B. W., Melgar, D., Bock, Y., Haase, J. S., & Geng, J. (2013). Earthquake magnitude scaling using seismogeodetic data. Geophysical Research Letters, 40(23), 6089–6094. https://doi.org/10.1002/2013GL058391

Crowell, B. W., Schmidt, D. A., Bodin, P., Vidale, J. E., Baker, B., Barrientos, S., & Geng, J. (2018). G-FAST earthquake early warning potential for great earthquakes in Chile. Seismological Research Letters, 89(2A), 542–556. https://doi.org/10.1785/0220170180

Crowell, B. W., Schmidt, D. A., Bodin, P., Vidale, J. E., Gomberg, J., Hartog, J. R., Kress, V. C., Melbourne, T. I., Santillan, M., Minson, S. E., & Jamison, D. G. (2016). Demonstration of the cascadia G-FAST geodetic earthquake early warning system for the Nisqually, Washington, Earthquake. Seismological Research Letters, 87(4), 930–943. https://doi.org/10.1785/0220150255

Dittmann, T., Liu, Y., Morton, Y., & Mencin, D. (2022). Supervised Machine Learning of High Rate GNSS Velocities for Earthquake Strong Motion Signals. Journal of Geophysical Research: Solid Earth, 127(11). https://doi.org/10.1029/2022JB024854

Fang, R., Shi, C., Song, W., Wang, G., & Liu, J. (2014). Determination of earthquake magnitude using GPS displacement waveforms from real-time precise point positioning. Geophysical Journal International, 196(1), 461–472. https://doi.org/10.1093/gji/ggt378

Ge, L., Han, S., Rizos, C., Ishikawa, Y., Hoshiba, M., Yoshida, Y., Izawa, M., Hashimoto, N., & Himori, S. (2000). GPS seismometers with up to 20 Hz sampling rate. In LETTER Earth Planets Space (Vol. 52).

Géron, A. (2019). Hands-On Machine Learning with Scikit-Learn, Keras, and TensorFlow: Concepts, Tools, and Techniques to Build Intelligent Systems (2nd ed.). O'Reilly Media.

Goldberg, D. E., Melgar, D., Hayes, G. P., Crowell, B. W., & Sahakian, V. J. (2021). A Ground-Motion Model for GNSS Peak Ground Displacement. Bulletin of the Seismological Society of America, 111(5), 2393–2407. https://doi.org/10.1785/0120210042

Goodfellow, I., Bengio, Y., & Courville, A. (2016). Deep Learning: Adaptive Computation and Machine Learning series. MIT Press. http://www.deeplearningbook.org

Grapenthin, R., Johanson, I. A., & Allen, R. M. (2014). Operational real-time GPS-enhanced earthquake early warning. Journal of Geophysical Research: Solid Earth, 119(10), 7944–7965. https://doi.org/10.1002/2014JB011400

Hanks, T. C., & Kanamori, H. (1979). A moment magnitude scale. Journal of Geophysical Research, 84(B5), 2348. https://doi.org/10.1029/JB084iB05p02348

Jiao, P., & Alavi, A. H. (2020). Artificial intelligence in seismology: Advent, performance and future trends. Geoscience Frontiers, 11(3), 739–744. https://doi.org/10.1016/j.gsf.2019.10.004

Kawamoto, S., Hiyama, Y., Ohta, Y., & Nishimura, T. (2016). First result from the GEONET real-time analysis system (REGARD): The case of the 2016 Kumamoto earthquakes. Earth, Planets and Space, 68(1). https://doi.org/10.1186/s40623-016-0564-4

Kingma, D. P., & Ba, J. (2015, December 22). Adam: A Method for Stochastic Optimization. Proceedings of the 3rd International Conference on Learning Representations (ICLR 2015). http://arxiv.org/abs/1412.6980

Kouba, J. (2003). Measuring Seismic Waves Induced by Large Earthquakes with GPS. In Stud. Geophys. Geod (Vol. 47). http://igscb.jpl.nasa.gov/organization/

Krizhevsky, A., Sutskever, I., & Hinton, G. E. (2017). ImageNet classification with deep convolutional neural networks. Communications of the ACM, 60(6), 84–90. https://doi.org/10.1145/3065386

Kuang, W., Yuan, C., & Zhang, J. (2021). Real-time determination of earthquake focal mechanism via deep learning. Nature Communications, 12(1). https://doi.org/10.1038/s41467-021-21670-x

Larson, K. M. (2009). GPS seismology. Journal of Geodesy, 83(3–4), 227–233. https://doi.org/10.1007/s00190-008-0233-x

Le Cun, Y. (1989). Generalization and network design strategies. In R. Pfeifer, Z. Schreter, F. Fogelman, & L. Steels (Eds.), *Connectionism in perspective* Elsevier.

Le Cun, Y., Bottou, L., Bengio, Y., & Haffner, P. (1998). Gradient-based learning applied to document recognition. Proceedings of the IEEE, 86(11), 2278–2324. https://doi.org/10.1109/5.726791

Li, W., Chakraborty, M., Fenner, D., Faber, J., Zhou, K., Rümpker, G., Stöcker, H., & Srivastava, N. (2022a). EPick: Attention-based multi-scale UNet for earthquake detection and seismic phase picking. Frontiers in Earth Science, 10. https://doi.org/10.3389/feart.2022.953007

Li, W., Chakraborty, M., Sha, Y., Zhou, K., Faber, J., Rümpker, G., Stöcker, H., & Srivastava, N. (2022b). A study on small magnitude seismic phase identification using 1D deep residual neural network. Artificial Intelligence in Geosciences, 3, 115–122. https://doi.org/10.1016/j.aiig.2022.10.002

Lin, J. -T., Melgar, D., Thomas, A. M., & Searcy, J. (2021). Early Warning for Great Earthquakes From Characterization of Crustal Deformation Patterns With Deep Learning. Journal of Geophysical Research: Solid Earth, 126(10). https://doi.org/10.1029/2021JB022703

Lin, J.-T., Melgar, D., Thomas, A., & Searcy, J. (2020). Chilean Subduction Zone rupture scenarios and waveform data. Zenodo. https://zenodo.org/record/5015610#.Y3to4S9t9pQ

Melgar, D., Crowell, B. W., Geng, J., Allen, R. M., Bock, Y., Riquelme, S., Hill, E. M., Protti, M., & Ganas, A. (2015). Earthquake magnitude calculation without saturation from the scaling of peak ground displacement. Geophysical Research Letters, 42(13), 5197–5205. https://doi.org/10.1002/2015GL064278

Melgar, D., LeVeque, R. J., Dreger, D. S., & Allen, R. M. (2016). Kinematic rupture scenarios and synthetic displacement data: An example application to the Cascadia subduction zone. Journal of Geophysical Research: Solid Earth, 121(9), 6658–6674. https://doi.org/10.1002/2016JB013314

Melgar, D., & Ruhl, C. (2018). High-rate GNSS displacement waveforms for large earthquakes version 2.0. Zenodo. https://doi.org/10.5281/zenodo.1434374

Mendoza, L., Kehm, A., Koppert, A., Martín Dávila, J., Gárate, J., & Becker, M. (2013). Un caso de estudio para la sismología GPS: el terremoto de Lorca. Física de La Tierra, 24(0). https://doi.org/10.5209/rev_fite.2012.v24.40135

Minson, S. E., Murray, J. R., Langbein, J. O., & Gomberg, J. S. (2014). Real-time inversions for finite fault slip models and rupture geometry based on high-rate GPS data. Journal of Geophysical Research: Solid Earth, 119(4), 3201–3231. https://doi.org/10.1002/2013JB010622

Mousavi, S. M., & Beroza, G. C. (2022). Deep-learning seismology. Science, 377(6607). https://doi.org/10.1126/science.abm4470

Nair, V., & Hinton, G. E. (2010). Rectified Linear Units Improve Restricted Boltzmann Machines. In J. Fürnkranz & T. Joachims (Eds.), ICML'10: Proceedings of the 27th International Conference on International Conference on Machine Learning (pp. 807–814). Omnipress.

Perol, T., Gharbi, M., & Denolle, M. (2018). Seismology Convolutional neural network for earthquake detection and location. https://www.science.org

Psimoulis, P. A., Houlié, N., Habboub, M., Michel, C., & Rothacher, M. (2018). Detection of ground motions using high-rate GPS time-series. Geophysical Journal International, 214(2), 1237–1251. https://doi.org/10.1093/gji/ggy198

Ruhl, C. J., Melgar, D., Chung, A. I., Grapenthin, R., & Allen, R. M. (2019). Quantifying the Value of Real-Time Geodetic Constraints for Earthquake Early Warning Using a Global Seismic and Geodetic Data Set. Journal of Geophysical Research: Solid Earth, 124(4), 3819–3837. https://doi.org/10.1029/2018JB016935

Ruhl, C. J., Melgar, D., Grapenthin, R., & Allen, R. M. (2017). The value of real-time GNSS to earthquake early warning. Geophysical Research Letters, 44(16), 8311–8319. https://doi.org/10.1002/2017GL074502

Satake, K. (2014). Advances in earthquake and tsunami sciences and disaster risk reduction since the 2004 Indian ocean tsunami. In Geoscience Letters (Vol. 1, Issue 1). SpringerOpen. https://doi.org/10.1186/s40562-014-0015-7

Scherer, D., Müller, A., & Behnke, S. (2010). Evaluation of Pooling Operations in Convolutional Architectures for Object Recognition. In K. Diamantaras & W. I. L. S. Duch (Eds.), Artificial Neural Networks – ICANN 2010. ICANN 2010. Lecture Notes in Computer Science (pp. 92–101). https://doi.org/10.1007/978-3-642-15825-4_10

Schmidhuber, J. (2015). Deep learning in neural networks: An overview. Neural Networks, 61, 85–117. https://doi.org/10.1016/j.neunet.2014.09.003

van den Ende, M. P. A., & Ampuero, J. P. (2020). Automated Seismic Source Characterization Using Deep Graph Neural Networks. Geophysical Research Letters, 47(17). https://doi.org/10.1029/2020GL088690

Wang, R., Parolai, S., Ge, M., Jin, M., Walter, T. R., & Zschau, J. (2013). The 2011 Mw 9.0 Tohoku earthquake: Comparison of GPS and strong-motion data. Bulletin of the Seismological Society of America, 103(2B), 1336–1347. https://doi.org/10.1785/0120110264

Zhou, & Chellappa. (1988). Computation of optical flow using a neural network. IEEE International Conference on Neural Networks, 71–78 vol.2. https://doi.org/10.1109/ICNN.1988.23914

# Appendix

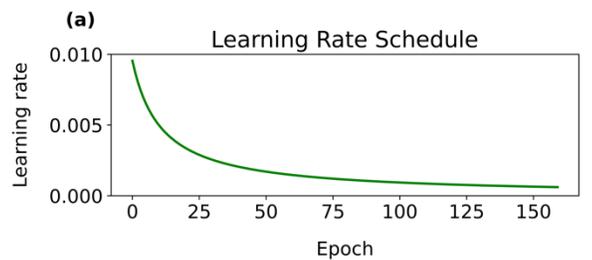

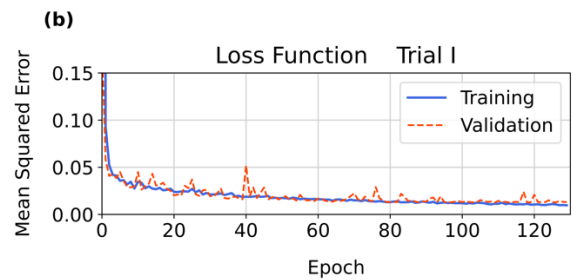

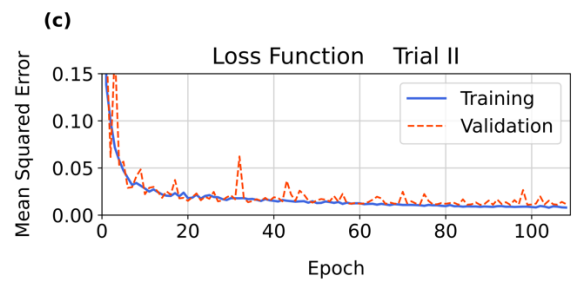

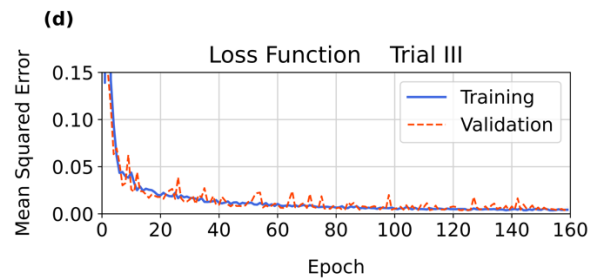

**Appendix A.** a) Learning rate function used during the training model. b), c), and d) show the model performance during the training and validation for every case described by Table 1. The loss values for both training and validation decrease, having a very slim gap between the curves, and reaching stability to the point of low MSE values: 0.02, 0.02, and 0.01, for Case I, Case II, and Case III, respectively. Also, the number of epochs necessary to get the minimum losses with a good fit of training and validation were: 139, 108, and 134 epochs, respectively.

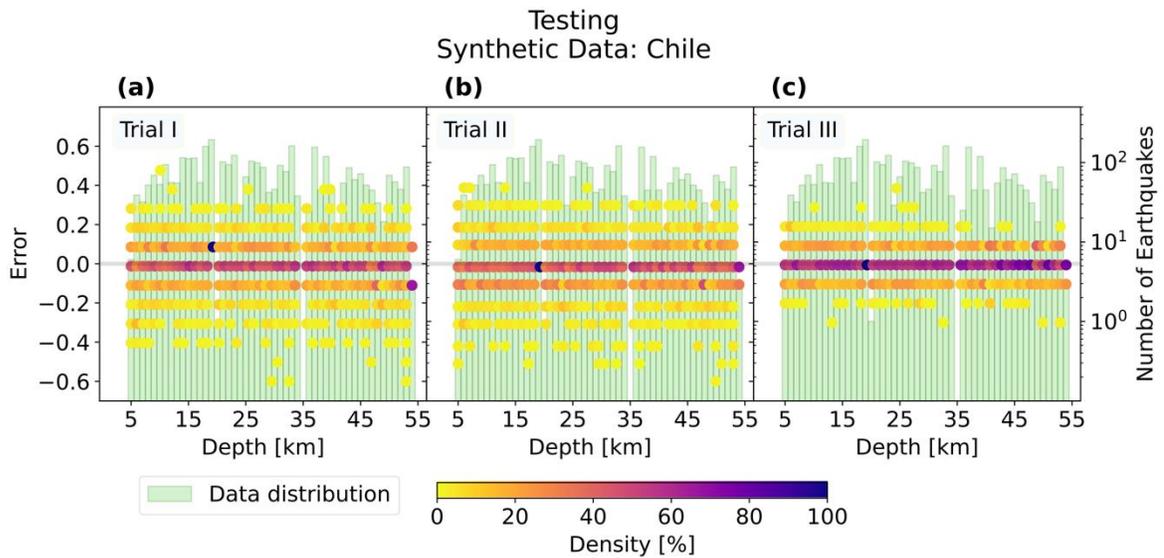

**Appendix B.** Error distribution by earthquake depth. The errors correspond to the difference between the real magnitude and the magnitude estimated by the models. The results are binned on a grid, 1 km × 0.1 magnitude error, and each circle corresponds to one bin whose color represents the percentage of tests done by each depth. The testing data distribution by depth is represented through histograms in the background of the plots. The error distributions suggest that there are no features in the time series related to the earthquake depth that could have an evident influence on the estimation.